\documentstyle[12pt]{article}
\newcommand{\journal}[4]{{\em #1~}#2\,(19#3)\,#4;}

\newcommand{\hpa}{\journal {Helv. Phys. Acta}}

\newcommand{\pr}{\journal {Phys. Rev.}}

\newcommand{\cmp}{\journal {Comm. Math. Phys.}}

\newcommand{\np}{\journal {Nucl. Phys.}}
\newcommand{\pl}{\journal {Phys. Lett.}}

\newcommand{\lmp}{\journal {Lett. Math. Phys.}}

\makeatletter
\@addtoreset{equation}{section}
\makeatother

\newcommand{\lp}{\left(}\newcommand{\rp}{\right)}

\renewcommand{\AA}{{\cal A}}
\newcommand{\BB}{{\cal B}}

\newcommand{\MM}{{\cal M}}

\newcommand{\SS}{{\cal S}}

\newcommand{\WW}{{\cal W}}

\newcommand{\bX}{{\bf X}}
\newcommand{\bY}{{\bf Y}}

\newcommand{\complex}{{\kern .1em {\raise .47ex
\hbox {$\scriptscriptstyle |$}}
    \kern -.4em {\rm C}}}
\newcommand{\real}{{{\rm I} \kern -.19em {\rm R}}}
\newcommand{\rational}{{\kern .1em {\raise .47ex
\hbox{$\scripscriptstyle |$}}
    \kern -.35em {\rm Q}}}
\renewcommand{\natural}{{\vrule height 1.6ex width
.05em depth 0ex \kern -.35em {\rm N}}}

\newcommand{\pa}{\partial}

\newcommand{\fud}[2]  {{\displaystyle{\frac{\delta #1}{\delta #2}}}}

\newcommand{\sla}{\raise.15ex\hbox{$/$}\kern -.57em}

\newcommand{\twiddle}{\lower.9ex\rlap{$\kern -.1em\scriptstyle\sim$}}

\newcommand{\equ}[1]{(\ref{#1})}

\newcommand{\eq}{\begin{equation}}
\newcommand{\eqn}[1]{\label{#1}\end{equation}}
\newcommand{\eea}{\end{eqnarray}}
\newcommand{\eqa}{\begin{eqnarray}}
\newcommand{\eqan}[1]{\label{#1}\end{eqnarray}}
\newcommand{\ba}{\begin{array}}
\newcommand{\ea}{\end{array}}
\newcommand{\eqac}{\begin{equation}\begin{array}{rcl}}
\newcommand{\eqacn}[1]{\end{array}\label{#1}\end{equation}}

\begin{document}
{\large     

{\ }

\vspace{20mm}
\centerline{\LARGE Supersymmetric structure of the bosonic string }
\vspace{2mm}

\centerline{\LARGE theory in the Beltrami parametrization}  \vspace{2mm}

\vspace{9mm}

\centerline{M. Werneck de Oliveira }
\centerline{{\small  International School for Advanced Studies} }
\centerline{{\small  Via Beirut 2-4, Trieste 34014, Italy }}
\vspace{4mm}
\centerline{ and }
\vspace{4mm}
\centerline{ M. Schweda, S.P. Sorella$^{1}$\footnotetext[1]{Supported in part
by the ''Fonds zur F\"orderung der Wissenschaftlichen Forschung'',
M008-Lise Meitner Fellowship.}  }
\centerline{{\small  Institut f\"ur Theoretische Physik} }
\centerline{{\small  Technische Universit\"at Wien }}
\centerline{{\small  Wiedner Hauptstra\ss e 8-10}}
\centerline{{\small  A-1040 Wien (Austria)}}
\vspace{4mm}
\vspace{10mm}

\centerline{{\normalsize {\bf REF. TUW 93-12 }} }

\vspace{4mm}
\vspace{10mm}

\centerline{\Large{\bf Abstract}}\vspace{2mm}
\noindent
We show that the bosonic string theory quantized in the Beltrami
parametrization possesses a supersymmetric structure like the
vector-supersymmetry
already observed in topological field theories.

\setcounter{page}{0}
\thispagestyle{empty}

\vfill
\pagebreak
\section{Introduction}

One of the most interesting features of the topological
models~\cite{Witten,Schwarz,BBRT} is represented by the existence of a
supersymmetric algebra whose generators describe the $BRS$ symmetry and the
vector supersymmetry carrying a Lorentz index~\cite{SUSY}.

Actually this supersymmetric structure turns out to be extremely useful
in discussing the renormalization of the topological models to all orders
of perturbation theory. Indeed, as shown in~\cite{GMS}, it provides an
elegant and simple way for solving the descent equations associated to the
integrated $BRS$ cohomology; yielding then a complete characterization of
all possible anomalies and invariant counterterms for both
Schwarz~\cite{LPS} and Witten's type~\cite{B} topological models.

The aim of this work is to show that this supersymmetric structure is
present also in the bosonic string quantized in a conformal gauge parametrized
by a Beltrami differential. This parametrization, introduced by~\cite{BBS},
allows to use a quantization procedure analogue to that of the
Yang-Mills theories. Moreover, as shown by~\cite{BECCHI}, the Beltrami
parametrization
turns out to be the most natural parametrization which exihibits the
holomorphic factorization of the Green functions according to the
Belavin-Polyakov-Zamolodchikov scheme~\cite{BPZ}.

The work is organized as follows: in Sect. 2 we briefly recall the $BRS$
quantization procedure; Sect. 3 is devoted to discuss the supersymmetric
algebra and the related Ward identities. Finally, in Sect. 4 we use the
aforementionned supersymmetry to solve the Wess-Zumino consistency
conditions for the Slavnov anomaly. In this letter we will limit
ourselves only to present the results without entering into technical
computations; a more complete and detailed version is in
preparation~\cite{PREP}.

\section{ Quantization and Slavnov identity}

Let us start with the bosonic string action
\eq
 S_{inv} = {1 \over 2} \int_{\MM} d^2 x \sqrt{g} g^{\alpha \beta}
                  \pa_{\alpha} \bf{X} \cdot \pa_{\beta} \bf{X}   \ ,
\eqn{sinv}
where $g_{\alpha\beta}$ $(\alpha,{\ }\beta=1,2)$ is a metric on the
two-dimensional string world sheet $\MM$ and $\{ \bX \}$ are the
string coordinates mapping $\MM$ into the $D$-dimensional real
plane $R^D$.

Denoting with
$(z,\bar z)$ a system of complex coordinates, the world sheet metric
$g$ can be parametrized by a Beltrami differential $\mu$~\cite{BBS,BECCHI}
\eq
  ds^2 =   g_{\alpha \beta} dx^{\alpha} dx^{\beta}
           \propto {\ } { \vert {\ } dz + \mu d{\bar z} {\ } \vert }^2  \ ,
\eqn{beltrami}
in terms of which the action \equ{sinv} takes the form
\eq
 S_{inv} = \int dz d \bar{z} { 1 \over {1-\mu \bar{\mu}}}
     \lp (1+ \mu \bar{\mu}) \pa \bf{X} \cdot \bar{\pa} \bf{X}
          - \mu \pa \bf{X} \cdot \pa \bf{X}
          -  \bar{\mu} \bar{\pa} \bf{X} \cdot \bar{\pa} \bf{X} \rp  \ ,
\eqn{smuinv}
with
\eq
   \pa = \pa_{z} \qquad \ , \qquad \bar{\pa} = \pa_{\bar z}  \ .
\eqn{ddbar}
As one can easily check $S_{inv}$ is invariant under an infinitesimal
diffeomorphism transformation generated by a two-components vector field
$(\gamma,\bar{\gamma})$:
\eq\ba{lcl}
     \delta \bf{X} =        \gamma \pa \bf{X}
                     + \bar{\gamma} \bar{\pa} \bf{X}  \ ,   \\
     \delta \mu   =  \gamma \pa \mu  + \bar{\gamma} \bar{\pa} \mu
                    + \mu \bar{\pa} \bar{\gamma}
                    + \bar{\pa} \gamma - \mu \pa \gamma
                    - \mu^2 \pa \bar{\gamma}  \ .
\ea\eqn{diff}
Following~\cite{BBS,BECCHI}, the quantization of $S_{inv}$ is done by
introducing the $(b,c)$ ghost system
\eq
 S_{bc} {\ }={\ } \int dz d \bar{z} {\ }b
                  \lp \bar{\pa} c + c\pa\mu -  \mu \pa c \rp  {\ }+ {\ }
                  \int dz d\bar{z} {\ }\bar{b}
                  \lp \pa \bar{c} + \bar{c} \bar{\pa} \bar{\mu}
                          - \bar{\mu} \bar{\pa} \bar{c} \rp  \ ,
\eqn{bcsystem}
so that the gauge fixed action
\eq
 S_{gf} =  S_{inv} + S_{bc} \ ,
\eqn{sgf}
is invariant under the nilpotent $BRS$ transformations~\cite{BBS,BECCHI}:
\eq\ba{lcl}
    s \bf{X} =  { {c -\mu \bar{c}} \over {1-\mu \bar{\mu}}}  \pa \bf{X}
                 {\ }+{\ }
                { {\bar{c} - \bar{\mu} c}\over {1-\mu \bar{\mu}}}
                 \bar{\pa} \bf{X}  \ ,   \\
    s c = c \pa c \ ,   \\
    s \bar{c} = \bar{c} \bar{\pa} \bar{c}  \ ,  \\
    s \mu = \bar{\pa}c + c \pa \mu - \mu \pa c  \ ,  \\
    s \bar{\mu} = \pa \bar{c} + \bar{c} \bar{\pa} \bar{\mu}
                              - \bar{\mu} \bar{\pa} \bar{c} \ ,  \\
    s b = s \bar{b} =  0  \ , \qquad     s^2 =0  \ .
\ea\eqn{brstr}
The ghosts $(c,\bar{c})$  in eqs.\equ{bcsystem}, \equ{brstr} have been
introduced by
C. Becchi~\cite{BECCHI} and are related to the diffeomorphism variables
$(\gamma, \bar{\gamma})$ of eq. \equ{diff} by
\eq
   c = \gamma + \mu \bar{\gamma} \ ,  \qquad
   \bar{c} = \bar{\gamma} + \bar{\mu} \gamma  \ .
\eqn{cgrel}
To translate the $BRS$ invariance of $S_{gf}$ into a Slavnov identity we
introduce a set of invariant external sources $(\bY, L, \bar{L})$ coupled
to the non-linear variations of the $BRS$ transformations of the
quantized fields \equ{brstr}:
\eq
   S_{ext} = \int dz d \bar{z} \lp \bY \cdot s {\bf{X}} + L c \pa c
                + \bar{L} \bar{c} \bar{\pa} \bar{c} \rp  \ .
\eqn{sext}
As explained in~\cite{BBS} $\mu$ and $\bar \mu$ are treated as external
unquantized fields playing the role of a background metric.

Thanks to the algebraic property
\eq
   s \mu{\ } = {\ }\fud{S_{gf}}{b}  \ ,
\eqn{muequat}
the complete action
\eq
  \Sigma{\ }={\ } S_{inv} + S_{bc} + S_{ext} \ ,
\eqn{totalaction}
obeys the Slavnov identity
\eq
  \SS(\Sigma) = 0 \ ,
\eqn{slavnov}
with
\eq
 \SS(\Sigma){\ }={\ } \int dz d \bar{z} \lp
           \fud{\Sigma}{\bY} \cdot \fud{\Sigma}{\bf{X}}
       +   \fud{\Sigma}{\mu} \fud{\Sigma}{b}
       +   \fud{\Sigma}{L} \fud{\Sigma}{c}
       +   \fud{\Sigma}{\bar{\mu}} \fud{\Sigma}{\bar{b}}
       +   \fud{\Sigma}{\bar{L}} \fud{\Sigma}{\bar{c}} \rp \ .
\eqn{slavexp}
Let us introduce also, for further use, the linearized Slavnov operator
$\BB$
\eq\ba{rl}
 \BB{\ }={\ } \int dz d \bar{z} ({\ }  &\!\!
           \fud{\Sigma}{\bY} \cdot \fud{\ }{\bf{X}}
       +   \fud{\Sigma}{\bf{X}} \cdot \fud{\ }{\bY}
       +   \fud{\Sigma}{\mu} \fud{\ }{b}
       +   \fud{\Sigma}{b} \fud{\ }{\mu}
       +   \fud{\Sigma}{L} \fud{\ }{c}  \\
 &\!\! +   \fud{\Sigma}{c} \fud{\ }{L}
       +   \fud{\Sigma}{\bar{\mu}} \fud{\ }{\bar{b}}
       +   \fud{\Sigma}{\bar{b}} \fud{\ }{\bar{\mu}}
       +   \fud{\Sigma}{\bar{L}} \fud{\ }{\bar{c}}
       +   \fud{\Sigma}{\bar{c}} \fud{\ }{\bar{L}} {\ }) \ ,
\ea\eqn{slavlin}
which, as a consequence of \equ{slavnov}, turns out to be nilpotent
\eq
  \BB \BB= 0  \ .
\eqn{nilp}
In this framework the Beltrami differential $\mu$ has a very simple physical
interpretation~\cite{BBS,BECCHI,Stora,Serge}: it is the classical source
for the $(T_{zz},{\ }T_{\bar{z} \bar{z}})$-components of the energy-momentum
tensor, i.e.:
\eq
  T_{zz} = \fud{\Sigma}{\mu}  \ , \qquad
  T_{\bar{z} \bar{z}} = \fud{\Sigma}{\bar{\mu}}  \ .
\eqn{energy}
The Slavonv identity \equ{slavnov} is then the starting point for a field
theory characterization of the energy-momentum current algebra~\cite{BPZ}.
{}From the expression \equ{slavlin} for the linearized operator $\BB$ it
follows also:
\eq
  T_{zz}{\ } ={\ } \BB b  \ ,
\eqn{benergy}
which shows that the energy-momentum tensor is cohomologically trivial. This
property, as discussed by~\cite{Witten}, is one of the basic ingredients for
the construction of topological field models.

Let us conclude this section by noticing that, actually, not only the
energy-momentum tensor but also the full string action $\Sigma$ in
\equ{totalaction} is cohomologically trivial. Indeed it is easily
verified that
\eq
  \Sigma{\ }={\ } \BB \int dz d \bar{z}
    \lp {1 \over 2} {\bf{X}} \cdot \bY - L c - \bar{L} \bar{c} \rp  \ .
\eqn{trivaction}
This property, together with eq.\equ{benergy}, allows to interpret in
a suggestive way the bosonic string as a topological model of the
Witten's type~\cite{Witten}

\section{ Supersymmetric algebra}

To discuss the symmetry content of the model and to show the existence
of a supersymmetric structure let us introduce
the two functional operators $(\WW,{\ }\bar{\WW})$:
\eq
  \WW{\ }={\ } \int dz d \bar{z}
         \lp \bar{\mu} \fud{\ }{\bar{c}} + \fud{\ }{c}
             + \bar{L} \fud{\ }{\bar b} \rp    \ ,
\eqn{woper}
\eq
 \bar{\WW}{\ }={\ } \int dz d \bar{z}
         \lp \mu \fud{\ }{c} + \fud{\ }{\bar c}
             + L \fud{\ }{b} \rp    \ ,
\eqn{bwoper}
which, together with the linearized Slavnov operator \equ{slavlin}, obey the
following algebraic relations
\eq
    \{ {\ }\BB,{\ }\WW{\ }\}= {\ }\pa   \ ,  \qquad
    \{ {\ }\BB,{\ }\bar{\WW}{\ }\}= {\ } \bar{\pa}    \ ,
\eqn{susyalg}
\eq
    \{ {\ }\WW,{\ }\WW{\ }\}= {\ }
    \{ {\ }\WW,{\ }\bar{\WW}{\ }\}={\ }
    \{ {\ }\bar{\WW},{\ }\bar{\WW}{\ }\}={\ }0   \ .
\eqn{susyalg1}
{}From eq.\equ{susyalg} one sees that the algebra between
$(\WW,{\ }\bar{\WW})$ and $\BB$ closes on the translations, thus allowing
for a supersymmetric interpretation of the model.

In addition, one has also the linearly broken Ward identities:
\eq
  \WW \Sigma{\ }={\ }\Delta \ ,  \qquad
  \bar{\WW} \Sigma{\ }={\ }\bar{\Delta}  \ ,
\eqn{susyward}
with  $(\Delta,{\ }\bar{\Delta})$ given by
\eq
 \Delta{\ }={\ }\int dz d \bar{z}
    \lp  \bar{L} \pa \bar{c} + L \pa c - \bar{b} \pa \bar{\mu}
         - b \pa \mu - \bY \cdot \pa \bf{X} \rp  \ ,
\eqn{delta}
and
\eq
 \bar{\Delta}{\ }={\ }\int dz d \bar{z}
    \lp  L \bar{\pa} c + \bar{L} \bar{\pa} \bar{c} - b \bar{\pa} \mu
         - \bar{b} \bar{\pa} \bar{\mu} - \bY \cdot \bar{\pa} \bf{X} \rp  \ .
\eqn{bdelta}
Expressions \equ{delta}-\equ{bdelta}, being linear in the quantum fields,
represent classical breakings. This property seems to be a common
feature of the models with a non-linearly
realized supersymmetry~\cite{PS}.

\section{ The diffeomorphism anomaly}

In this section we use the supersymmetric structure \equ{susyalg}
in order to solve the descent equations associated with the integrated
cohomology of the linearized operator $\BB$; giving then an algebraic
characterization of all possible anomalies of the Slavnov
identity \equ{slavnov} at the quantum level.

In what follows we identify, for simplicity, the string world sheet
$\MM$ with the whole
complex plane $C$; the result being adaptable, modulo the infrared
problem of the global zero modes~\cite{Stora}, to an arbitrary Riemann
surface by means of a projective connection~\cite{Serge,Zucchini}.

To the quantum level the classical action \equ{totalaction} is replaced
by a one-loop effective action $\Gamma$~\cite{BBS,BECCHI,Stora}
\eq
   \Gamma{\ }={\ }\Sigma + \hbar \Gamma^{(1)}  \ ,
\eqn{oneloop}
which obeys the anomalous Slavnov identity
\eq
  \SS(\Gamma){\ }={\ }\hbar \AA  \ ,
\eqn{anomaly}
where the diffeomorphism anomaly $\AA$ is an integrated  local two-form
of ghost number one \footnote{ We adopt here the usual convention of
denoting with $\AA_{q}^{p}$ a $q$-form with ghost number equal to $p$. }:
\eq
  \AA{\ }={\ }\int \AA^{1}_{2}  \ ,
\eqn{intanom}
constrained by the Wess-Zumino consistency condition
\eq
  \BB \AA = 0  \ .
\eqn{wesszumino}
As it is well known this equation, when translated to the local level,
yields a tower of descent equations:
\eq\ba{lcl}
          \BB \AA^1_2{\ }+{\ }d \AA^2_1{\ }={\ }0  \ ,  \\
          \BB \AA^2_1{\ }+{\ }d \AA^3_0{\ }={\ }0  \ ,  \\
          \BB \AA^3_0                  {\ }={\ }0  \ ,
\ea\eqn{desceq}
where $d$ denotes the exterior derivative
\eq
    d{\ }={\ }dz \pa + d\bar{z} \bar{\pa}    \ ,
\eqn{extderiv}
and
\eq
   d^2=0 \qquad \ ,  \qquad  \{ \BB{\ },{\ }d \} = 0  \ .
\eqn{bbdalg}
Thanks to the supersymmetric operators $(\WW,{\ }\bar{\WW})$, to
solve the ladder \equ{desceq} it is sufficient to know~\cite{PREP} only
the non-trivial solution of the last equation ( which is a problem of local
cohomology instead of a modulo-$d$ one ). It is easy to check indeed that,
once $\AA^3_0$ is known, the remanent cocycles $\AA^2_1$ and $\AA^1_2$ are
identified with the $(\WW,{\ }\bar{\WW})$-transform of $\AA^3_0$, i.e.
\eq\ba{lcl}
   \AA^2_1{\ }={\ }\lp \WW \AA^3_0 \rp dz {\ }+{\ }
                   \lp \bar{\WW} \AA^3_0 \rp d\bar{z} \ ,  \\
   \AA^1_2{\ }={\ }\lp \WW  \bar{\WW} \AA^3_0 \rp
                    dz \wedge d\bar{z} \ .
\ea\eqn{solut}
It is worthwhile to mention also that, due to the
vanishing of the local cohomology of $\BB$ in the one-form sector with
ghost number two and in the two-form sector with ghost number
one~\cite{PREP,BANDELLONI},  expression \equ{solut} is, modulo trivial
cocycles, the most general solution of the ladder \equ{desceq}.

For what concerns the local cohomology of $\BB$ in the zero-form sector
with ghost number three it turns out~\cite{PREP,BANDELLONI} that the most
general non-trivial solution for $\AA^3_0$ contains only two elements:
\eq
   \AA^3_0{\ }={\ }\lp \Omega^{(1)},{\ }\Omega^{(2)} \rp  \ ,
\eqn{omegasol}
which, modulo a $\BB$-coboundary, can be written as:
\eq
   \Omega^{(1)}{\ }={\ }\lp c \pa c \pa^2 c
          {\ }+{\ } {comp.}{\ }{\ }{conj.}  \rp  \ ,
\eqn{omega1}
and
\eq
   \Omega^{(2)}{\ }={\ } \lp c \bar{c} \pa c
              (D{\bf{X}} \cdot \bar{D} {\bf{X}})  f({\bf{X}})
               {\ }+{\ } {comp.}{\ }{\ }{conj.}   \rp  \ ,
\eqn{omega2}
where $D$ is the "covariant derivative"~\cite{Stora,Serge}
\eq
       D{\ }={\ } {1 \over {1 -\mu \bar{\mu}}}
                   (\pa - {\bar \mu} \bar{\pa}) \ ,
\eqn{covder}
and $f(\bf{X})$ is an arbitrary formal power series in the matter fields
$\{ \bf{X} \}$ which does not contain constant term:
\eq
    f({\bf{X}}){\ }={\ }\sum_{n=1}^{\infty} f_n
                {  \lp {\bf{X}} \cdot {\bf{X}} \rp  }^n  \ .
\eqn{fexpress}
Using equations \equ{solut} it is immediate to show that to the cocycle
$\Omega^{(1)}$ of eq.\equ{omega1} it corresponds the
diffeomorphism anomaly
\eq
  \AA^1_2{\ }={\ }
         \lp - \pa\mu \pa^2 c + \pa c \pa^2\mu \rp dz \wedge d\bar{z}
        {\ }+{\ }{comp.}{\ }{\ }{conj.} \ ,
\eqn{aa12}
i.e.
\eq
  \AA{\ }\propto \int dz d\bar{z}
      \lp \mu \pa^3 c {\ }+{\ }  {comp.}{\ }{\ }{conj.}  \rp  \ .
\eqn{aa}
This anomaly, whose numerical coefficient turns out to be proportional
to $(D-26)$~\cite{BECCHI}, corresponds to the well known
central term of the energy-momentum current algebra~\cite{BPZ} and fixes
the critical dimensions of the bosonic string.
Moreover it is completely equivalent~\cite{ST}, via a Bardeen-Zumino
action, to the more popular string Weyl anomaly.

The $\Omega^{(2)}$-cocycle of eq.\equ{omega2} gives rise to a matter
dependent cocycle $\AA_{\bf X}$ whose expression reads:
\eq
  \AA_{\bf X}{\ }={\ }-  \int dz d\bar{z} (1-\mu \bar{\mu})
           \lp (\pa + \bar{\mu} \bar{\pa} )\gamma +
               (\bar{\pa} + \mu \pa )\bar{\gamma} \rp
              (D{\bf{X}} \cdot \bar{D} {\bf{X}})  f(\bf{X})   \ ,
\eqn{aanomx}
where
\eq
  \gamma{\ }={\ } { {c - \mu \bar{c}} \over {1-\mu \bar{\mu}}} \ ,
\eqn{invg}
is the diffeomorphism variable of eqs.\equ{diff}, \equ{cgrel}. Actually to
$\AA_{\bf X}$ one cannot associate a true anomaly. Indeed, from the absence
in the classical action \equ{totalaction} of a self-interaction term in the
matter fields, it is easily seen that, in spite of the fact that $\AA_{\bf X}$
is cohomologically non-trivial, the numerical coefficient of the corresponding
Feynman diagrams automatically vanishes. It follows then that
expression \equ{aa} represents the unique breaking of the Slavonv
identity \equ{slavnov} at the quantum level.

{\ }

\noindent{\large{\bf Acknowledgements}}
We are grateful to O. Piguet, S. Lazzarini and R. Stora for useful
discussions and comments.

{\ }


\end{document}